\documentclass[11pt,twocolumn,twoside,a4paper,amsmath,amssymb,aps,showkeys,showpacs]{revtex4}

\usepackage{amsfonts}
\usepackage{graphics} %   <------- for LatTeX + EPS
\usepackage{epsfig}
\usepackage{fancyheadings}

\begin{document}
\thispagestyle{myheadings}
%%%%%%%%%%%%%%%%%%%%%%%%%% Title %%%%%%%%%%%%%%%%%%%%%%%%%%%%%%%%%%%%%%
\rhead[]{}%<------
\lhead[]{}%<------
\chead[Y. Hama, R.P.G. Andrade, F. Grassi, W.-L. Qian] 
{Ridge Effect in Hydrodynamics}%<------short title

\title{Trying to understand the ridge effect in hydrodynamic model}

\author{Yogiro Hama}
\email{hama@fma.if.usp.br}

\author{Rone Peterson G. Andrade}
%\email{randrade@fma.if.usp.br}

\author{Fr\'ed\'erique Grassi}
%\email{grassi@fma.if.usp.br}

\author{Wei-Liang Qian}
%\email{wlqian@fma.if.usp.br}

\affiliation{%
Instituto de F\'{\i}sica, Universidade de S\~ao Paulo,
C. Postal 66318, 05314-970 S\~ao Paulo,  BRAZIL }%

\received{ ?????????? }

\begin{abstract}
In a recent paper, 
%\cite{jun},  
the hydrodynamic code NeXSPheRIO was used in conjunction with STAR analysis methods to study two-particle correlations as function of $\Delta\eta$ and $\Delta\phi$. Both the ridge-like near-side and 
the double-hump away-side structures were obtained. 
However, the mechanism of ridge production was not clear. In order to understand it, we study a simple model with only one high-energy density peripheral tube in a smooth cylindrical back-ground, with longitudinal boost invariance. The results are rather surprising, but the model does produce the triple-ridge structure with one high ridge plus two lower ones placed symmetrically with respect to the former one. 
The shape of this structure is rather stable in a wide range of parameters. 

\end{abstract}



\maketitle

\renewcommand{\thefootnote}{\fnsymbol{footnote}}
{\footnotetext[1]{Also delivered by YH as a talk at WPCF2009
Workshop at CERN, Geneva (Switzerland), October 2009}

\renewcommand{\thefootnote}{\roman{footnote}}

%*****************   The Body of the Article:   *************************

\section{Introduction}
\label{introduction}

The {\it Ridge Effect} has been observed in two-particle long-range correlation measurements 
\cite{star1,phenix1,phobos}. The main characteristic is a 
narrow $\Delta\phi$ and wide $\Delta\eta$ correlation around the trigger. There is also some awayside stucture but experimentally less known, especially with respect to $\Delta\eta$. Originally, the trigger was chosen a high-$p_T$, presumably jet, particle, but now data are available also for low-$p_T$  trigger or no-trigger \cite{phenix2}. 
  
In a previous work \cite{jun}, we got the ridge structure in 
a hydrodynamic model. In hydrodynamic approach of nuclear  collisions, it is assumed that, after a complex process  involving microscopic collisions of nuclear 
constituents, at a certain early instant a hot and dense matter is formed, which would be in local thermal  equilibrium. This state is characterized by some initial conditions (IC), usually parametrized as smooth distributions of thermodynamic quantities and four-velocity. However, evidently smooth IC do not produce ridges. Since 
our systems are small, important event-by-event fluctuations  are expected in real collisions. In earlier works, by using  NeXuS event generator \cite{nexus}, we introduced fluctuating 
IC in hydrodynamics and studied several effects. 
In particular, the fluctuations of $v_2$, showing good 
agreement with data \cite{osada,qm2002,kiev}. We call 
NeXSPheRIO the junction of NeXuS with our hydro code SPheRIO  
\cite{sph}, based on the smoothed-particle hydrodynamic 
algorithm. The IC for high-energy nuclear collisions are not  
only event-by-event fluctuating but, if the thermalization 
is verified at very early time, they should be very 
bumpy. In Figure$\,$\ref{IC}, we show an example of such IC, 
generated by NeXuS, for a Au+Au collision at 200A GeV. 
Observe the tubular structure in $\eta$. 
  
The main objects of this contribution are {\it i}) to show some further results with our NeXSPheRIO code, which will be done in the next Section; and {\it ii}) to discuss what we learned about the mechanism of ridge production in our hydro model. The latter will be done in Section$\,$\ref{origin}. 

\section{Some Results on Two-Particle Correlation}
\label{Some Results}

In this Section, we are going to show some of our recent results on two-particle correlation in Au+Au collisions at 200A GeV. 

\begin{figure*} 
\hspace{-1.cm} 
\begin{center}
\includegraphics[width=8.5cm]{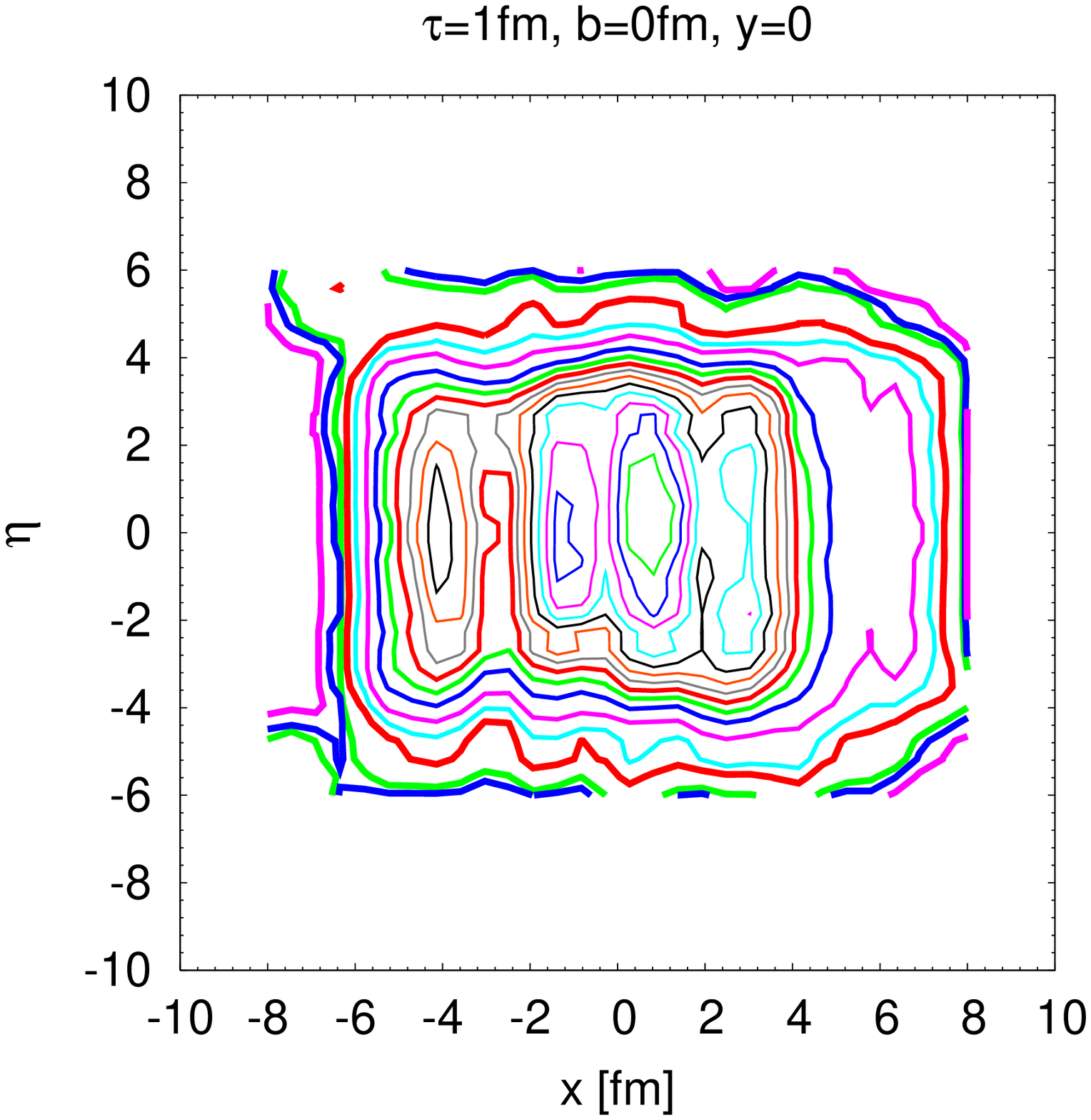} 
\hspace{-1.5cm} 
\includegraphics[width=8.5cm]{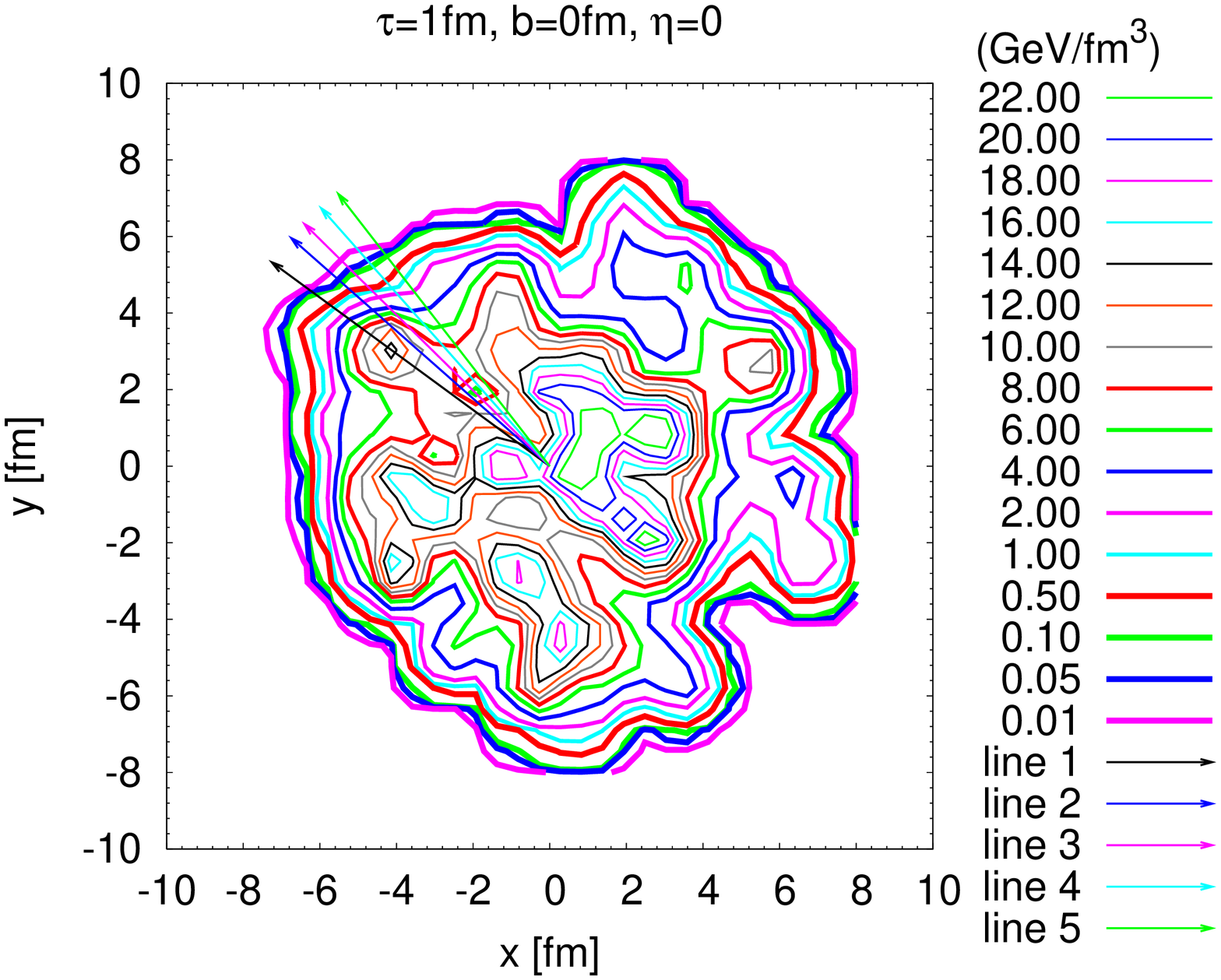}  
\caption{\label{IC}Energy density distribution at $\tau=1$fm for a central Au+Au collision at $200A$ GeV, given by NeXuS generator, in the collision plane (left) and in the 
mid-rapidity transverse plane (right).} 
\end{center}
\end{figure*}
\medskip

\begin{figure*} 
\vspace{2.cm} 
\includegraphics[width=12.cm]{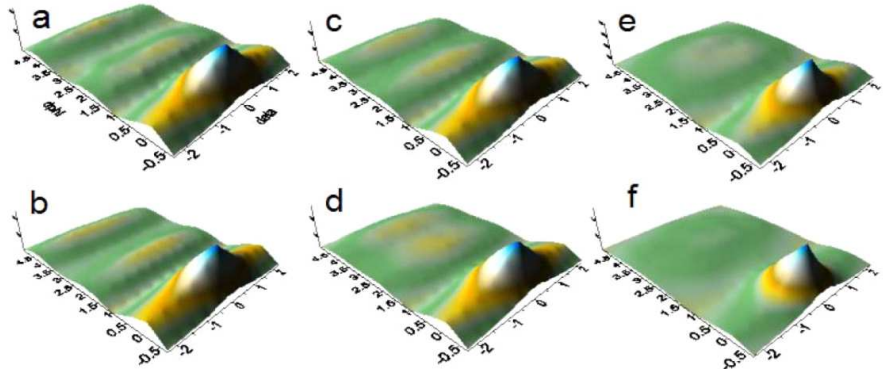} 
\vspace{-5.5cm} 
\caption{\label{CentDep}Two-particle correlation, as computed with our NeXSPheRIO code, for different centrality windows of Au+Au collisions at 200A GeV(a: 0-5\%, b: 5-10\%, c: 10-20\%, d: 20-30\%, e: 30-40\%, 
f: 40-50\%). 
The transverse 
momenta are chosen as $p_T^{trig}>2.5\,$GeV and $p_T^{ass}>1.5\,$GeV 
for triggers and associated particles, respectively.} 
\end{figure*}

\subsection{Centrality dependence} 

Data on the centrality dependence of the two-particle correlation have been reported
%in Ref.$\,$
\cite{phenix2}. 
In general, the nearside ridge decreases in height with decreasing centrality, or going from most central to peripheral windows, and at the same time the awayside structure in $\Delta\phi$ changes from double humps to single 
peak. In Figure$\,$\ref{CentDep}, we show our results with %the 
NeXSPheRIO. 
%code. 
It is clearly seen there the tendency we described above (compare with Figures 36-38 of Reference  \cite{phenix2}). 

\subsection{In-plane out-of-plane effect} 

Another effect which has been experimentally studied is 
the dependence of the correlation on the azimuthal angle 
$\phi_t$ of the trigger with respect to the event 
plane \cite{star2}. In a mid-central window, the away-side 
structure in $\Delta\phi$ is a peak at $\Delta\phi=\pi$, 
if the trigger is close to the event plane, and it is split 
into two peaks, as $\phi_t$ goes closer to $\pi/2$. 
We show in Figure$\,$\ref{inout} our results for the 
20 - 30\% centrality window. It is seen that the results 
show precisely the expected behavior from the data (see 
Figure$\,$1 of Reference$\,$\cite{star2}). 
\medskip 

\begin{figure}
\hspace*{-1.2cm} 
\includegraphics[angle=90,width=10.5cm]{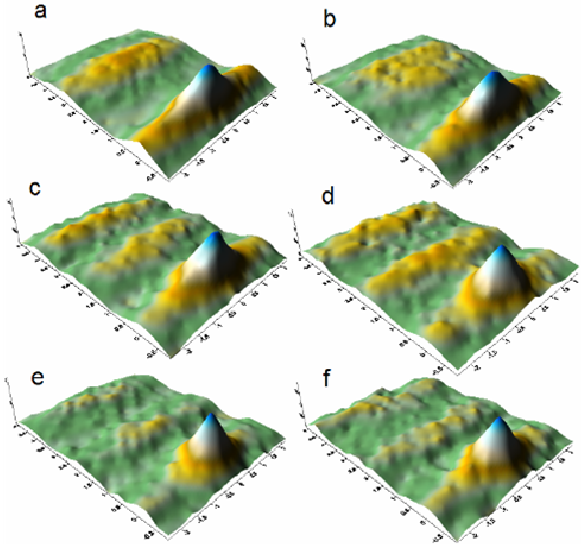} 
\vspace*{-.6cm} 
\caption{\label{inout}Two-particle correlation for 20-30\% 
centrality Au+Au collisions at 200A GeV, for different  %trigger-particle azimuthal angles 
$\phi_t$ 
(a: $\phi_t=0^\circ-15^\circ$, b: $\phi_t=15^\circ-30^\circ$, c: $\phi_t=30^\circ-45^\circ$, d: $\phi_t=45^\circ-60^\circ$, e: $\phi_t=60^\circ-75^\circ$, f: $\phi_t=75^\circ-90^\circ$).
%Transverse momenta have been chosen 
$p_T^{trig}>3.\,$GeV and $p_T^{ass}>1.\,$GeV.} 
\vspace*{-.5cm} 
\end{figure}

\section{Origin of the ridge-structure in hydrodynamics} 
\label{origin}

As seen, the NeXSPheRIO code does produce nice results on 
ridges, if we restrict ourselves to appropriate $p_T$ domain 
for hydrodynamic model application. However, {\it what is the origin of ridges?} Since each event in our model presents IC with many high-energy density tubes, one may associate these tubes $+$ transverse expansion with the ridge structure. But, 
the phenomenon is not so trivial. Moreover, {\it why 
away-side ridges?} By considering mainly the central 
collisions, we tried to understand the origin of the ridge 
structure, especially the away-side one. 

\subsection{Method of study - 2D model} 

Let us fix our attention to one of the tubes, located close 
to the surface of the hot matter, for instance the one where the arrow 1 passes on in Figure$\,$\ref{IC}. To study closely 
what happens in the neighborhood of this tube, we replace 
the complex background, as shown there, by a smooth one. 
Then, we use a 2D model with  boost-invariant longitudinal 
expansion, to simplify the computation. We find these 
assumptions quite reasonable for our purpose. 

Then, we parametrize the energy density as  
\begin{equation}
  \epsilon=12\exp[-.0004r^5]+\frac{34}{.845\pi} 
  \exp[-\frac{|{\bf r}-{\bf r}_0|^2}{.845}]\ ,  
  \label{par} 
\end{equation} 
where $r_0=5.4\,$fm, 
and the initial velocity of the fluid as 
\begin{equation}
  v_T=\tanh[4.57\exp(-27.2/r)]\ , 
  \label{parv}
\end{equation} 
corresponding to the average NeXuS IC. 

We show in Figure$\,$\ref{lines}, a comparison of the 
parametrization above with the original energy density  distribution as given by the NeXuS event we are studying. 
Notice that, except for the inner region, the agreement 
is reasonable. 

\begin{figure}[h]
\vspace*{.45cm} 
\includegraphics[width=7.1cm]{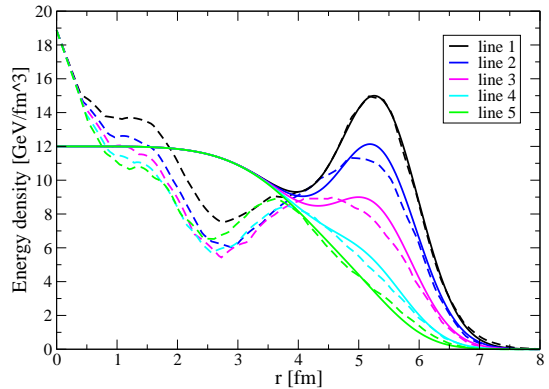}
\vspace*{-.15cm} 
\caption{\label{lines}Comparison of the parametrization 
given by Eq.(\ref{par}) (solid lines) with the original 
energy density, as given in Figure$\,$\ref{IC} (dashed 
lines), along the lines 1-5.} 
\vspace*{-.3cm} 
\end{figure}

\subsection{Results} 

What does the high-energy tube produce, in conjunction with the background? The result is: it deflects the otherwise isotropic radial flow of the background, in such a way to produce two symmetrical peaks in the resultant flow. 
In Figure~\ref{phi-distributions}, we show the azimuthal  distribution of the produced particles in different $p_T$ 
intervals, with respect to the tube position, set 
$\phi_{tube}=0$. 

\begin{figure*} 
  \begin{center} 
%   \vspace{-2.5cm} 
%   $\frac{dN}{d\phi}$
%   \hspace*{-.3cm} 
%   \vspace{2.5cm}
   \includegraphics[width=5.4cm]{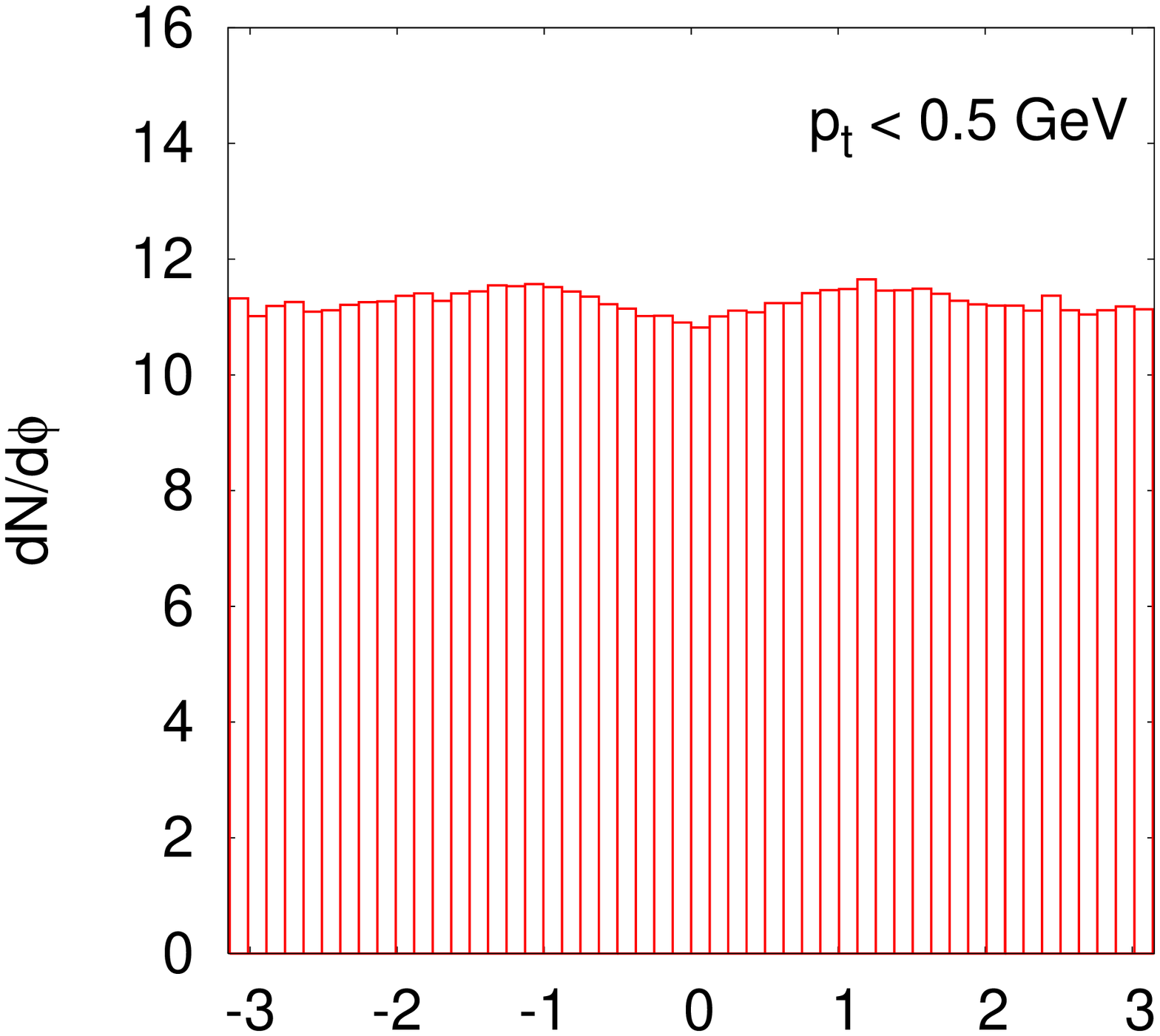} 
   \hspace*{-2.cm}
%   \includegraphics[width=5.4cm]{05_pt_10.eps} 
%   \hspace*{-2.cm}    
   \includegraphics[width=5.4cm]{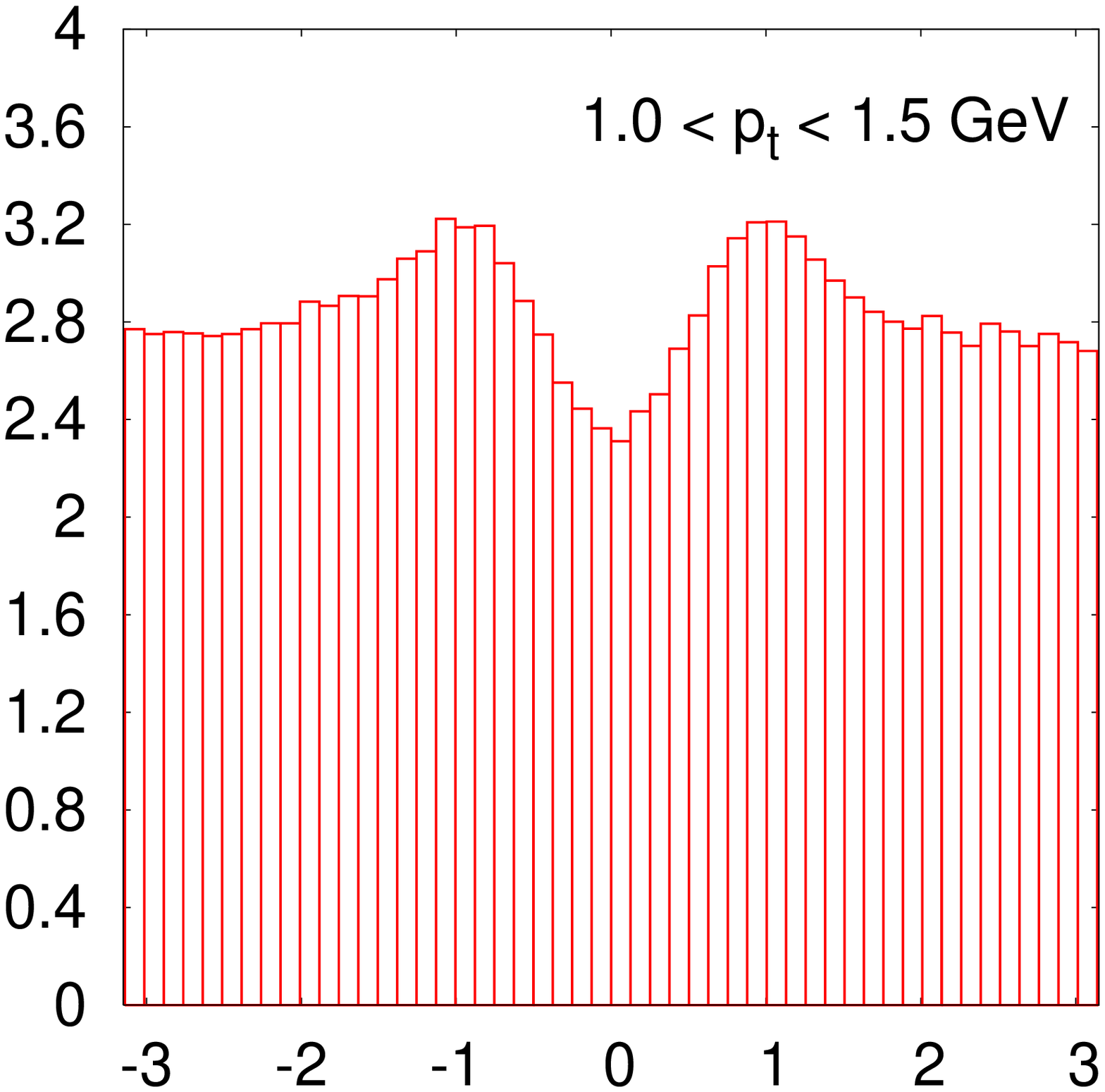}
   \hspace*{-2.cm} 
%   \includegraphics[width=5.4cm]{15_pt_20.eps}\\ 
%   \hspace*{.2cm} 
   \includegraphics[width=5.4cm]{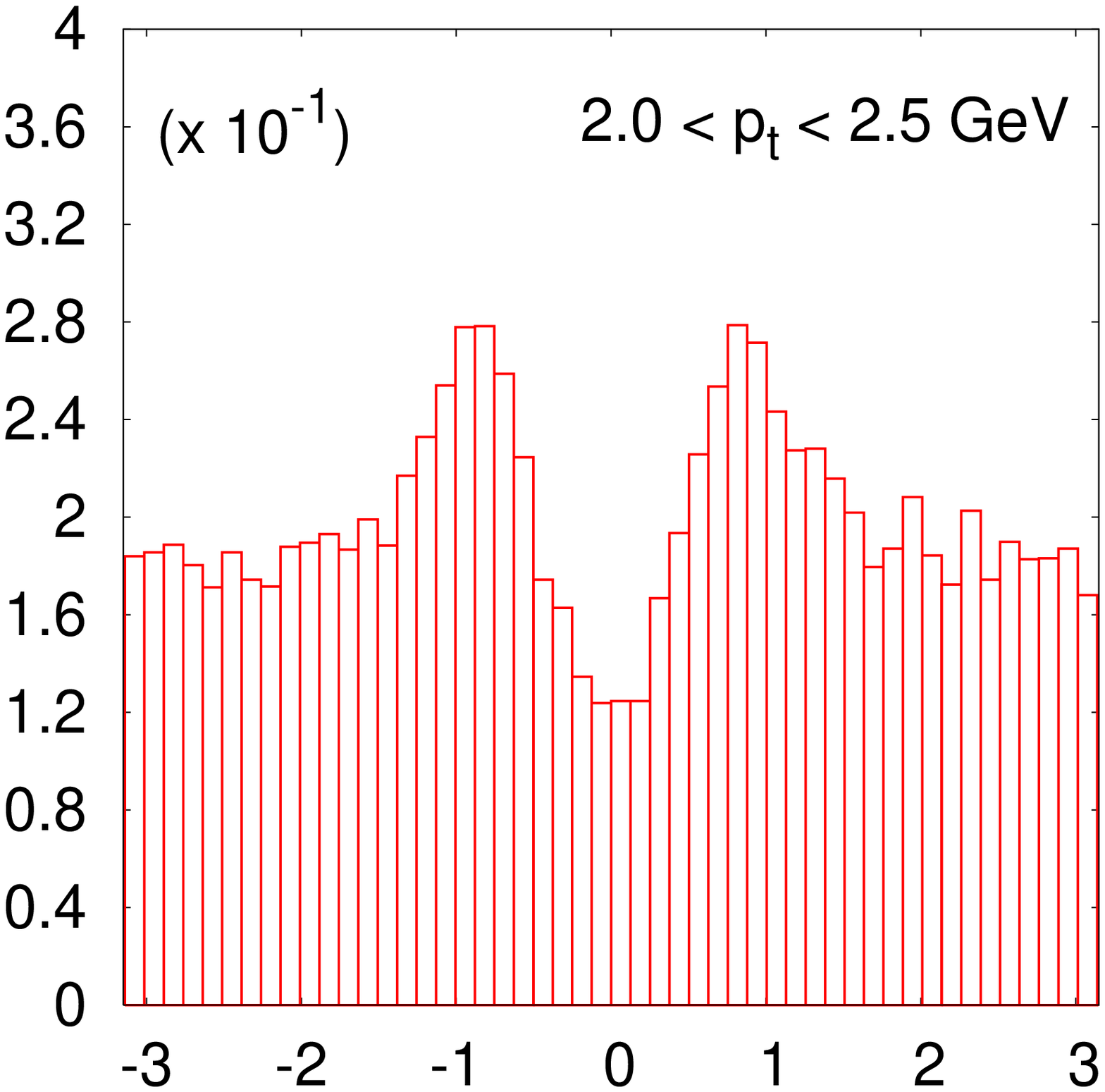}
   \hspace*{-2.cm}    
%   \includegraphics[width=5.4cm]{25_pt_30.eps}
%   \hspace*{-2.cm}    
   \includegraphics[width=5.4cm]{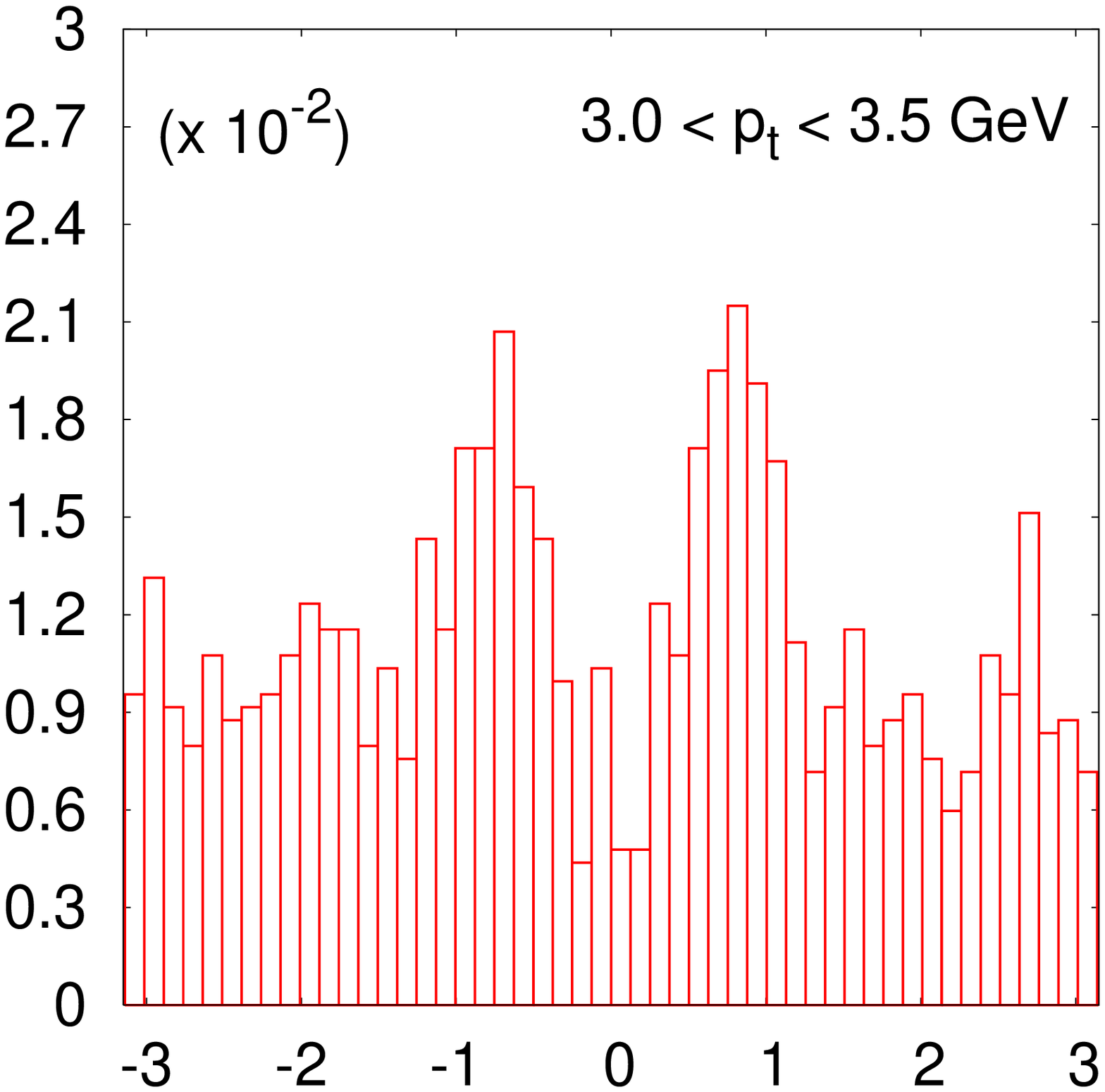}
   \hspace*{-2.cm}    
   \\ 
   $\phi$
  \end{center}  
  \vspace*{-.5cm}  
  \caption{\label{phi-distributions}Angular distributions of 
  particles in some different $p_T$ intervals, in the 2D model.} 
\end{figure*}

\begin{figure}%[h]
%\vspace*{-.2cm} 
\includegraphics[width=5.4cm]{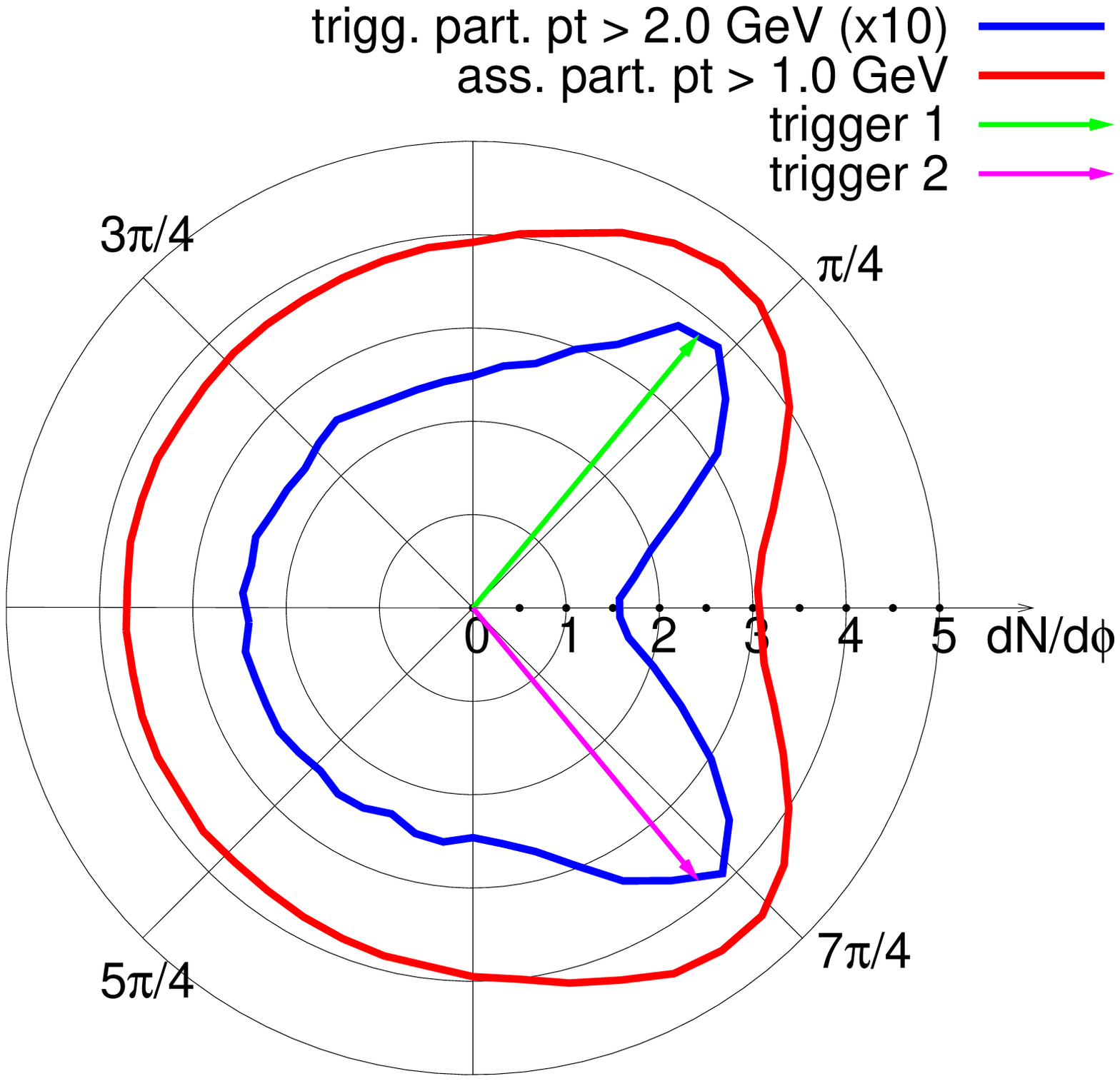} 
\vspace*{-.5cm} 
\includegraphics[width=5.3cm]{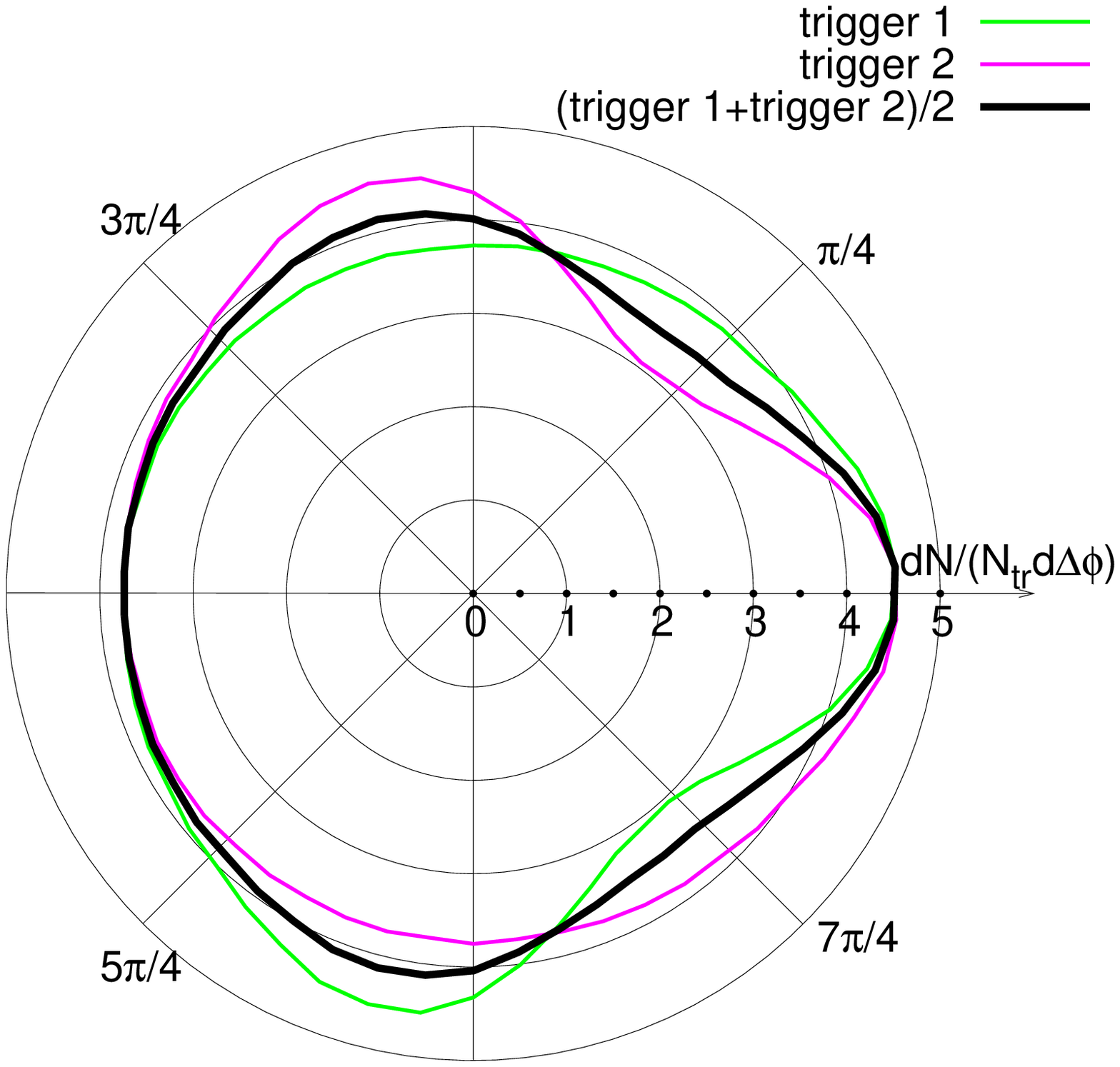} 
%\vspace*{-.5cm} 
%\includegraphics[width=4.5cm]{corrt4.eps} 
%\includegraphics[angle=90,width=6.5cm]{corr2.eps} 
\caption{\label{polar}Top panel: Angular distributions of trigger particles ($p_T>2$GeV) and the associated particles
($p_T>1$GeV). Bottom panel: Corresponding correlations, if 
the triggers are 1) at $\sim\pi/4$ (green); 2) at $\sim-\pi/4$ 
(cian); and 3) half of them at $\sim\pi/4$ and the other half at $\sim-\pi/4$ (black).} 
\end{figure}

Let us see what the corresponding two-particle correlation  
looks like. In Figure$\,$\ref{polar}, we show similar plots 
as in Figure$\,$\ref{phi-distributions}, but in polar 
coordinates and for triggers with $p_T>2$GeV and associate 
particles with $p_T>1$GeV. 
Now, if a trigger is found at $\phi_t\sim\pi/4$ (trigger 1), 
then, the corresponding correlation will be exactly the associated-particle distribution plotted there, but turned 
clockwise the same angle $\phi_t\,$. Similarly, if a trigger 
is at $\phi_t\sim-\pi/4$ (trigger 2), the correlation will be 
equal to the associated-particle distribution, but now turned 
couter-clockwise the same angle. Now, the triggers 1 and 2 
are equally probable, so for a sample of events with half of 
the triggers of type 1 and the other half of type 2, the 
correlation will be an average of these two as plotted 
in black in the bottom panel of Figure$\,$\ref{polar}. 
Actually, we have to integrate over $\phi_t$ to obtain the 
resultant correlation in our model. The results are shown in 
Figure$\,$\ref{pt_dependence}, together with the 
associated-particle-momentum dependence.  

\begin{figure}%[h]
\vspace*{-.2cm} 
\includegraphics[width=5.cm]{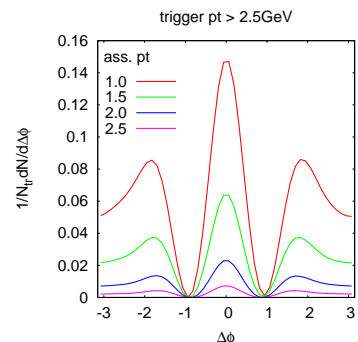} 
%\vspace*{-.5cm} 
%\includegraphics[angle=90,width=7.cm]{corr2.eps} 
\caption{\label{pt_dependence}Associated-particle $p_T$ 
dependence of the two-particle correlation.} 
\end{figure}

\begin{figure*} 
 \vspace*{-.3cm}
 \begin{center} 
  \includegraphics[width=4.3cm]{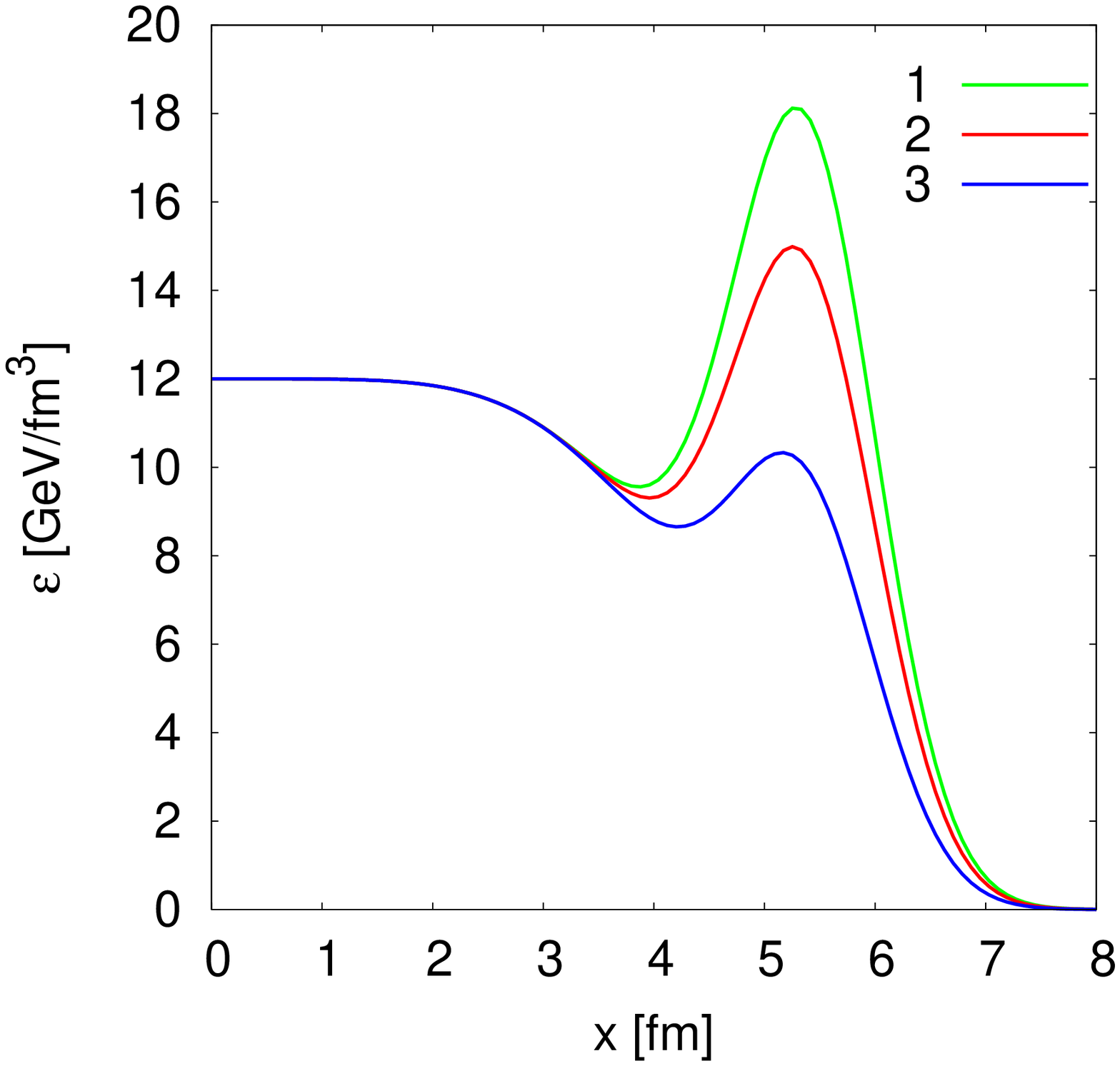} 
  \includegraphics[width=4.3cm]{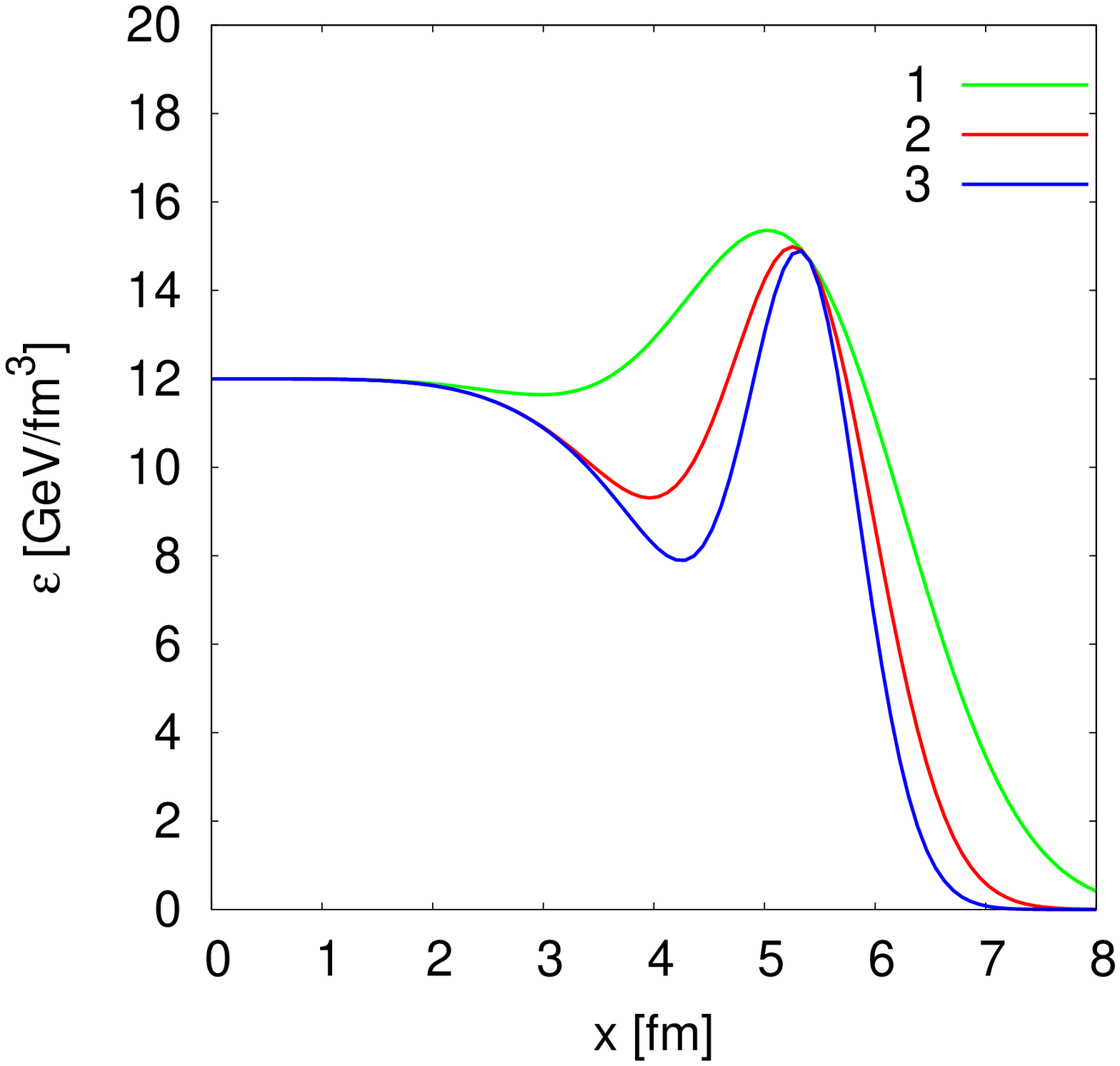} 
  \includegraphics[width=4.3cm]{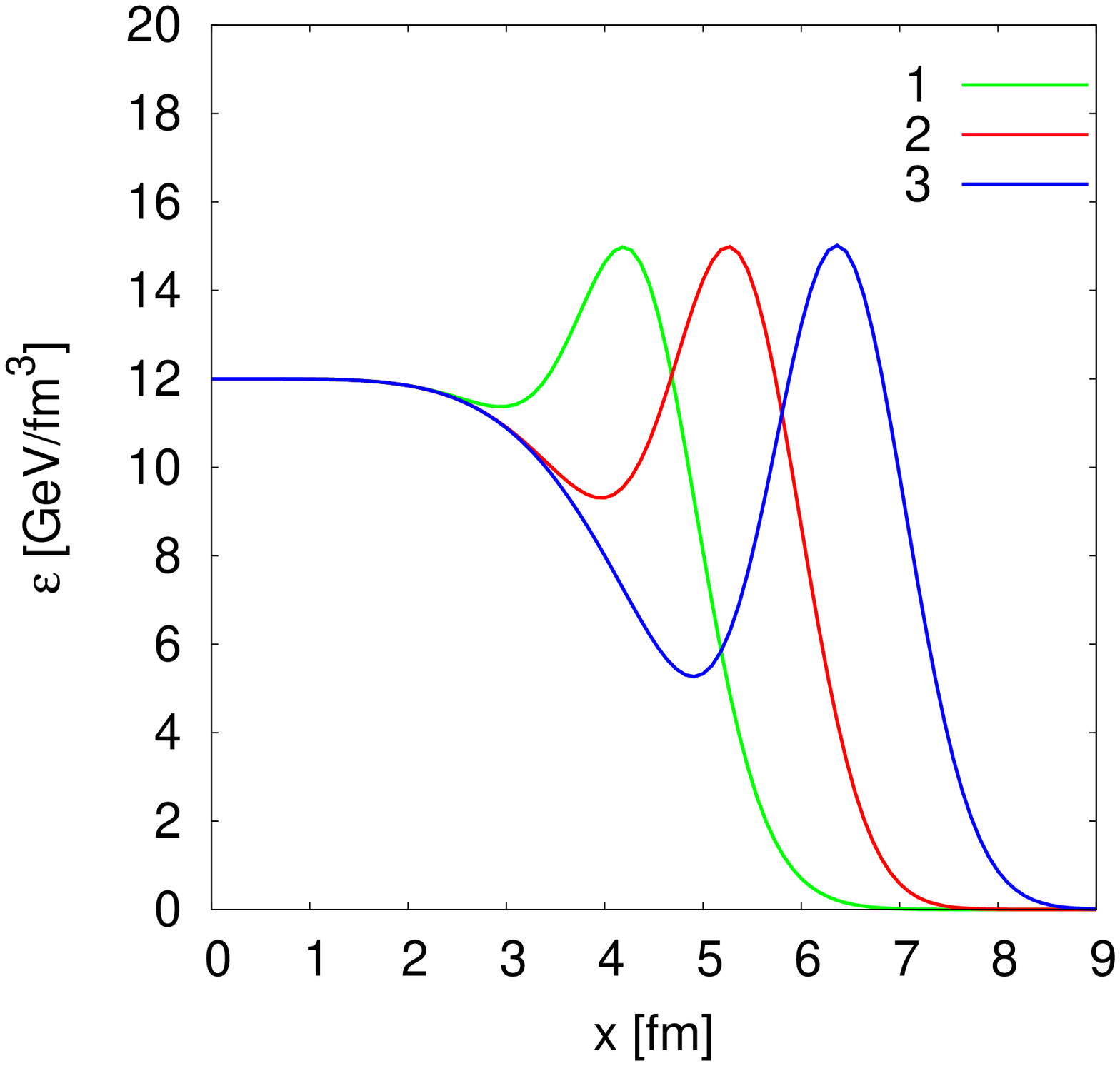}\\ 
  \includegraphics[width=4.3cm]{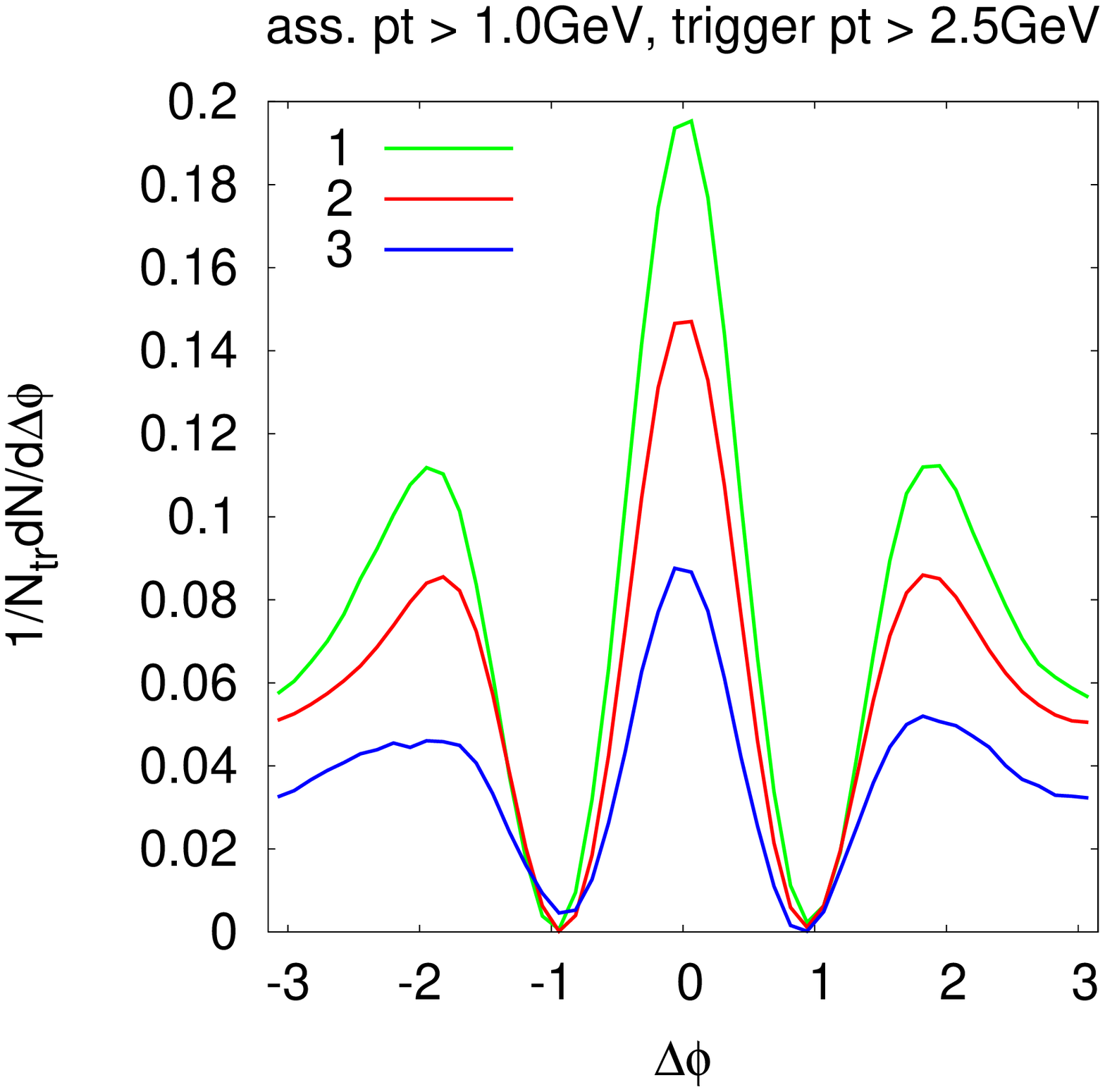}
  \includegraphics[width=4.3cm]{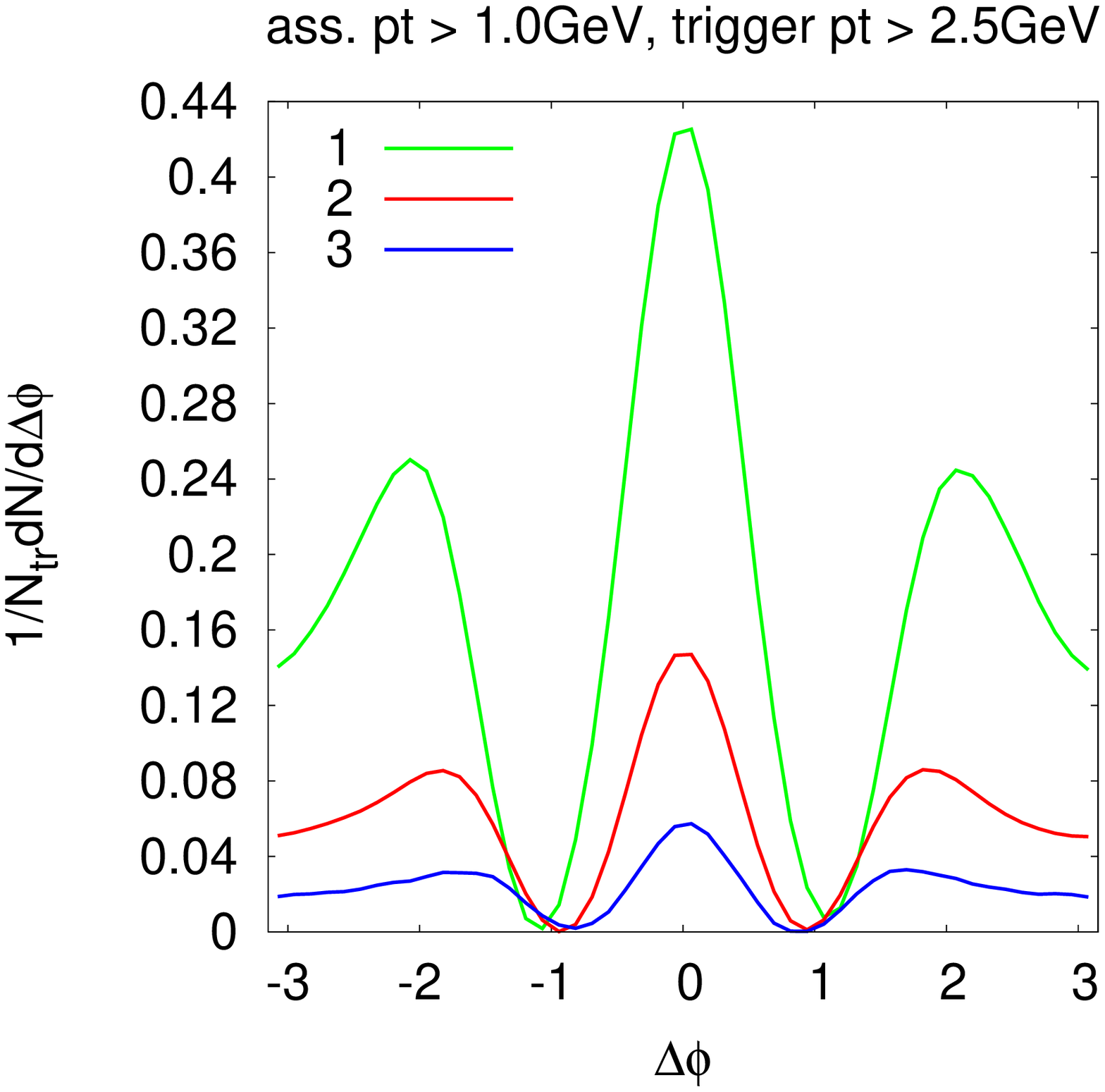}
  \includegraphics[width=4.3cm]{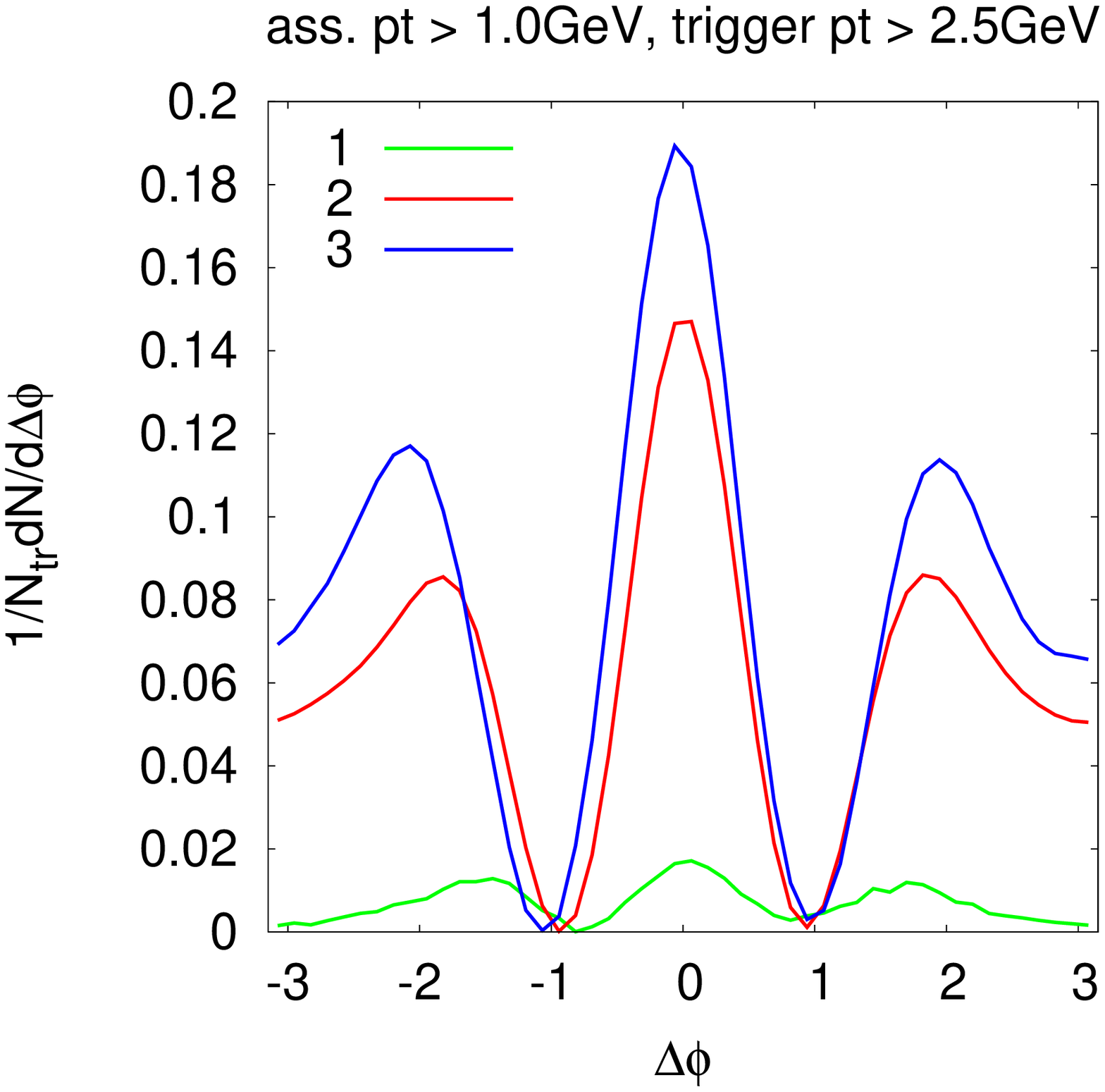}
 \end{center}  
  \vspace*{-.8cm}  
  \caption{\label{par_dependence}Parameter dependence of the 
  two-particle correlation in our model. Height (1st. column), radius (2nd. column) and position (3rd. column) dependences, respectively.} 
  \vspace*{-.2cm} 
\end{figure*}

\subsection{Parameter dependence} 

We studied how our previous results depend on the several 
parameters which define our energy density and the velocity 
distributions of IC given by Eqs.~(\ref{par}) and 
(\ref{parv}). The two-particle correlation is almost 
insensitive to some of the parameters like the height of 
the background and the transverse velocity. Other parameters 
importantly affect the intensity of the correlation, without 
changing the shape of the three-peak structure. See some of 
the results in Figure$\,$\ref{par_dependence}. 
An important result, which we could not include here due to the space limitation, is the shape of the background. Long-tailed background like Gaussian one, does not produce strong correlation through the mechanism described here. 

\section{Conclusions} 
\label{conclusions} 
In conclusion, the hydrodynamic expansion starting from 
fluctuating IC with tubular structure produces the ridge 
structure in the 2-particle correlation. We showed in this 
paper that the NeXSPheRIO code can reproduce several observed 
characteristics of the ridge effect. 

In central collisions, a high-density tube, close to the 
surface of the hot matter, causes flow with two maxima in 
azimuth, symmetrical with respect to the tube position. 
Such a flow implies a near-side peak and double, symmetrical 
away-side peaks in $\Delta\phi$ in the 2-particle 
correlation, with respect to the high-$p_T$ trigger. 

\newpage     
The shape of 2-particle correlation curve is more or less 
stable in a wide range of parameters. The intensity of the 
correlation depends strongly on the height and the radius of 
the tube and its position, but not sensitive neither to the 
height of the background nor to the initial transverse 
velocity. The shape of the background is an important 
factor. Long-tailed background like Gaussian one, does not
produce strong correlation through this mechanism. 
\smallskip

%\centerline{\bf Acknowledgments} 
%\bigskip 
\noindent{\bf Acknowledgments:} 
This work received support from Funda\c c\~ao de Amparo \`a 
Pesquisa do Estado de S\~ao Paulo, 
FAPESP, 
%under the contract nos.$\,$2009/50180-0 and 2005/54595-9, 
and Conselho Nacional de Desenvolvimento Cient\'{\i}fico e Tecnol\'ogico, CNPq.

\label{last}
\end{document}